\documentclass[aps,prl,twocolumn,superscriptaddress,longbibliography,nobibnotes]{revtex4-2}

\usepackage{diagbox}
\usepackage{amsmath}
\usepackage{amsfonts}
\usepackage{physics}
\usepackage[space]{grffile}
\usepackage{graphicx}
\usepackage{amssymb}
\usepackage{upgreek}
\usepackage{lineno}
\usepackage{natbib}
\usepackage{braket}
\usepackage{float}
\usepackage{color}
\usepackage{blindtext}
\usepackage{comment}

\setcitestyle{super}
\usepackage{xr-hyper}
\usepackage{hyperref}
\usepackage{xcolor}
\hypersetup{colorlinks,breaklinks,
    urlcolor=[rgb]{0,0,0.64},
    linkcolor=[rgb]{0,0,0.64},
    citecolor=[rgb]{0,0,0.64},
    filecolor=[rgb]{0,0,0.64}}

\usepackage[T1]{fontenc}

\newcommand{\MIT}{Massachusetts Institute of Technology, Department of Physics, Cambridge, Massachusetts 02139, USA.}
\newcommand{\StanfordAP}{Department of Applied Physics, Stanford University, Stanford, California 94305, USA.}

\newcommand{\SIMES}{SIMES, SLAC National Accelerator Laboratory, Menlo Park, California 94025, USA.}
\newcommand{\GLAM}{Geballe Laboratory for Advanced Materials, Stanford University, Stanford, California 94305, USA.}
\newcommand{\ETH}{Institute for Theoretical Physics, ETH Z{\"u}rich, 8093 Z{\"u}rich, Switzerland.}
\newcommand{\MITEECS}{Massachusetts Institute of Technology, Department of Electrical Engineering and Computer Science, Cambridge, Massachusetts 02139, USA.}
\newcommand{\Harvard}{Lyman Laboratory, Department of Physics, Harvard University, Cambridge, Massachusetts 02138, USA.}
\newcommand{\UTokyo}{Department of Applied Physics, University of Tokyo, Bunkyo-ku, Tokyo 113-8656, Japan.}
\newcommand{\TDL}{Tsung-Dao Lee Institute, School of Physics and Astronomy, and Zhangjiang Institute for Advanced Study, Shanghai Jiao Tong University, Shanghai 200240, China.}
\newcommand{\StanfordPAP}{Departments of Physics and of Applied Physics, Stanford University, Stanford, California 94305, USA.}

\begin{document}

\title{Time-domain identification of distinct mechanisms for competing charge density waves in a rare-earth tritelluride}

\author{Yifan~Su}
\thanks{These authors contributed equally to this work: Y.S., B.-Q.L., and A.Z.}
\affiliation{\MIT}
\author{B.~Q.~Lv}
\thanks{These authors contributed equally to this work: Y.S., B.-Q.L., and A.Z.}
\affiliation{\TDL}
\affiliation{\MIT}
\author{Alfred~Zong}
\thanks{These authors contributed equally to this work: Y.S., B.-Q.L., and A.Z.}
\affiliation{\MIT}
\affiliation{\StanfordPAP}
\affiliation{\SIMES}
\author{Aaron~M{\"u}ller}
\affiliation{\ETH}
\author{Sambuddha~Chattopadhyay}
\affiliation{\ETH}
\affiliation{\Harvard}
\author{Pavel~E.~Dolgirev}
\affiliation{\Harvard}
\author{Anisha~G.~Singh}
\affiliation{\SIMES}
\affiliation{\GLAM}
\affiliation{\StanfordAP}
\author{Joshua~A.~W.~Straquadine}
\affiliation{\SIMES}
\affiliation{\GLAM}
\affiliation{\StanfordAP}
\author{Dongsung~Choi}
\affiliation{\MITEECS}
\author{Doron~Azoury}
\affiliation{\MIT}
\author{Masataka~Mogi}
\affiliation{\MIT}
\affiliation{\UTokyo}
\author{Ian~R.~Fisher}
\affiliation{\SIMES}
\affiliation{\GLAM}
\affiliation{\StanfordAP}
\author{Eugene~Demler}
\affiliation{\ETH}
\author{Nuh~Gedik}
\email[Correspondence to: ]{gedik@mit.edu}
\affiliation{\MIT}

\begin{abstract}
    Understanding the origin of phase transitions and the interactions between distinct phases remains a central task in condensed matter physics. Charge density wave (CDW) systems provide an ideal platform for investigating these phenomena. While the dominant CDW phases in many materials can be explained through Fermi surface nesting or electron-phonon interactions, certain CDW phase transitions remain poorly understood, challenging conventional paradigms. One notable example is rare-earth tritelluride ErTe$_3$, which hosts two competing CDW orders. While the dominant CDW phase fits within the electron-phonon coupling framework, the formation mechanism of the subdominant CDW remains enigmatic. In this study, we combine time- and angle-resolved photoemission spectroscopy and time-dependent Ginzburg-Landau theory to establish a time-domain approach for probing phase transitions in solid-state systems. By analyzing the distinct recovery dynamics of the two CDW orders in ErTe$_3$ following light excitation, we reveal a novel nucleation and growth mechanism that likely drives the secondary CDW phase transition. This work not only uncovers a previously unknown CDW formation mechanism in rare-earth tritellurides but also introduces a non-equilibrium framework for understanding phase transitions and phase competition in quantum materials.
\end{abstract}

\date{\today}

\maketitle

In modern statistical mechanics, phase transitions are generally classified as either first-order or second-order, based on the continuity properties of the free energy and its derivatives\cite{Goldenfeld2018}. Despite decades of research, the experimental classification and underlying mechanisms of phase transitions in many physical systems remain elusive. A major challenge lies in the difficulty of directly probing the free energy landscape in equilibrium, which governs these transitions. An alternative approach is to drive the system out of equilibrium using an ultrafast laser pulse that transiently weakens or quenches the ordered phase. As the system relaxes back to equilibrium along with its free energy surface, tracking the temporal evolution of relevant observables can provide insights into both the underlying free energy landscape and the mechanisms driving the phase transition\cite{Zong2019,Kogar2020,Zong2019b,Zong2021RoleOrder,Wandel2022,Nova2019,Liu2024,delaTorre2021Colloquium:Materials}.

In this study, we apply this non-equilibrium approach to a charge density wave (CDW) transition. CDWs are characterized by periodic modulations of charge density accompanied by lattice distortions\cite{Gruner2018DensitySolids}. The order parameter of a CDW can be directly measured using photoemission spectroscopy (via the energy gap in the electronic density of states) or diffraction (via the intensity of CDW satellite peaks)\cite{Gruner2018DensitySolids,Zong2019b,Warren1990X-rayDiffraction,Sobota2021Angle-resolvedMaterials}. This makes them ideal platforms for studying phase transition mechanisms. While conventional CDW transitions are typically driven by electron-phonon interactions, many CDW systems deviate from this framework\cite{Zhu2015ClassificationNature,Johannes2008FermiMetals}. By employing time-resolved techniques, we aim to uncover unconventional CDW formation mechanisms and develop a more comprehensive understanding of the microscopic processes governing phase transitions in quantum materials.

\begin{figure*}[tb!]
\centering
\includegraphics[width=1\textwidth]{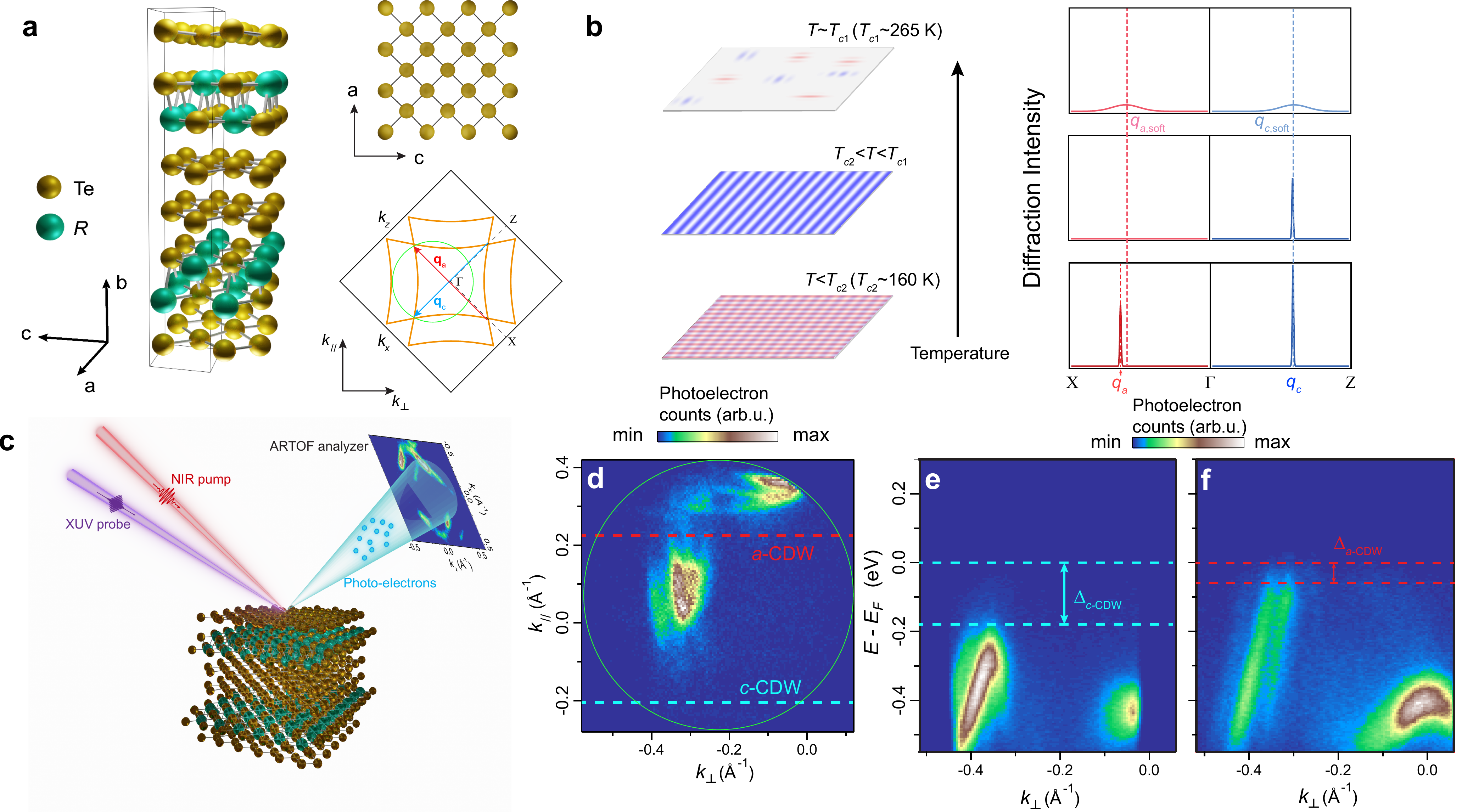}
\caption{\textbf{Charge density wave transitions in ErTe$_\text{3}$ observed with equilibrium ARPES.} \textbf{a}, Left: Schematic of the ErTe$_3$ crystal structure, where the black lines indicate the unit cell. Top-right: Schematic of Te sheet, which is nearly square-shaped with a slight in-plane anisotropy. Bottom-right: Schematic of the normal state Fermi surface arising from the planar Te sheets. The blue and red arrows indicate the wavevectors $\mathbf{q}_c$ and $\mathbf{q}_a$ of the equilibrium $c$-axis and $a$-axis CDW. The green circle indicates the portion of the Fermi surface accessed by 10.8~eV ARPES. \textbf{b} Schematics of CDW structures in ErTe$_3$ and corresponding diffraction superlattice peaks. Top: near $T_{c1}$, phonons soften at $q_\text{soft}$ along both the $a$-axis and $c$-axis, resulting in short-range charge density fluctuations, manifested as weak broad peaks along both axes in the diffraction pattern. Middle: Between $T_{c1}$ and $T_{c2}$, long range $c$-CDW forms along $c$-axis. The phonon at $q_\text{soft}$ along $a$-axis start to harden back. Bottom: below $T_{c2}$, long range order forms along $a$-axis, at a momentum different from $q_\text{soft}$. \textbf{c}, Schematic of the time- and angle-resolved photoemission spectroscopy (trARPES) experiment. ARTOF stands for the angle-resolved time-of-flight analyzer for photoelectrons. \textbf{d} Fermi surface map of the ARPES spectrum, taken at 45~K ($T<T_{c2}$). The green circle corresponds to the one in \textbf{a}. Following the convention in previous literature, $k_\perp$ and $k_\parallel$ are oriented at 45$^\circ$ with respect to the in-plane axes $a$ and $c$. $k_{\parallel}\equiv(k_z+k_x)/\sqrt{2}$, $k_{\perp}\equiv(k_z-k_x)/\sqrt{2}$. \textbf{e,f} Electronic dispersion along $k_\perp$ at corresponding cuts in \textbf{d}, indicated by color-coded dashed lines. The corresponding CDW gaps, $\Delta_{c\text{-CDW}}$ and $\Delta_{a\text{-CDW}}$, are labeled. }
\label{fig:Fig1}
\end{figure*}

\begin{figure*}[htb!]
\centering
\includegraphics[width=0.75\textwidth]{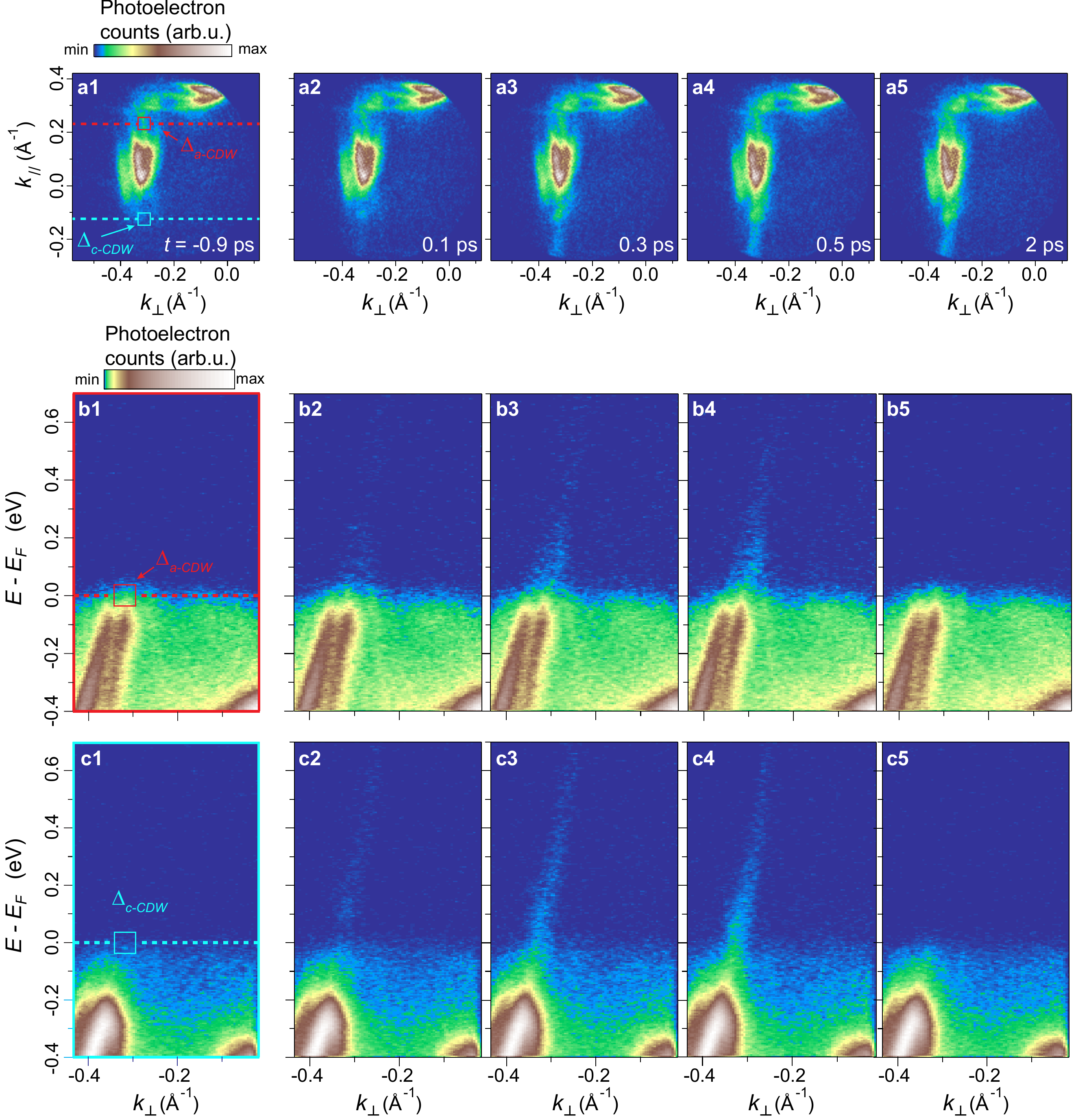}
\caption{\textbf{Snapshots of CDW gap dynamics upon photoexcitation.} \textbf{a1--a5} Time evolution of the Fermi surface after photoexcitation. In a1, the blue and red boxes highlight the gaps opened by $c$- and $a$-CDW respectively. \textbf{b1--b5, c1--c5} Time evolution of band structure near the Fermi surface after photoexcitation. The dispersion cuts in \textbf{b} and \textbf{c} are respectively taken along red and blue dashed lines in \textbf{a1}. The red (blue) box in \textbf{a1} and \textbf{b1} (\textbf{c1}) defines the three-dimensional region of interest.
}
\label{fig:Fig2}
\end{figure*}

A prominent example of CDW with unidentified origin is those in rare-earth tritelluride ($R$Te$_3$, where $R$ denotes a rare-earth element), a family of prototypical CDW materials. $R$Te$_3$ possess a layered, quasi-tetragonal structure [Fig.~\ref{fig:Fig1}a] with a slight in-plane anisotropy ($a/c\geq0.997$ at 300~K)\cite{Malliakas2006} leading to a preferred direction of the CDW along the $c$-axis. This dominant CDW is hereafter referred to as $c$-CDW. The Fermi surface, which is similar for all $R$Te$_3$, arises from the Te bilayer square-net sheets \cite{Yumigeta2021AdvancesSynthesis}. The normal-state Fermi surface calculated from the tight-binding model is depicted in the lower-right corner of (Fig.~\ref{fig:Fig1}a). The rare-earth element can be used as a tuning knob to adjust the chemical pressure and thus the ground state of the material. Materials with light rare-earth atoms only undergo one CDW transition, forming a $c$-CDW below the transition temperature ($T_{c1}$). In materials with rare-earth atoms heavier than Gd, a second CDW order, typically referred to as $a$-CDW, emerges along the $a$-axis, perpendicular to the $c$-axis in the same Te sheet, at a lower transition temperature ($T_{c2}$)(See Fig.~\ref{fig:Fig1}b, detailed discussion on the nomenclature can be found in Supplemental Information). Though sharing almost the same structural motifs and nesting vectors on the Fermi surface, the $a$-CDW is not simply the $c$-CDW rotated by 90$^\circ$. One prominent difference is that the $a$-CDW is not associated with a soft phonon mode around $T_{c2}$ from inelastic X-ray measurements\cite{Maschek2018Competing/math} (see Fig.~\ref{fig:Fig1}b), arguing against a stereotypical second-order CDW phase transition \cite{Gruner2018DensitySolids,Varma1983Strong-CouplingTransitions}. On the other hand, a recent Raman scattering study on $R$Te$_3$ observed amplitude mode softening towards transition temperature (ref.~\cite{Singh2024}), implicitly contradicting the previous X-ray scattering and optical measurements \cite{Maschek2018Competing/math,Lavagnini2010,Lavagnini2008,Yumigeta2022,Pfuner2010}. Moreover, recent transport measurements discovered a strong violation of Wiedemann-Franz law below $T_{c2}$, indicating the breakdown of typical electron-phonon coupling picture of CDW formation \cite{Kountz2021AnomalousTransition}. These findings point toward an exotic origin of the $a$-CDW that motivates the current study. 

In the past decade, multiple attempts were made to understand the $a$-CDW phase transition through measurements of the CDW order parameters. Direct characterization of CDW order parameters can be accomplished in two ways. From an electronic perspective, the amplitude of the CDW order parameter can be read out from the electronic gap opened near the Fermi surface \cite{Moore2010Fermi/math,Gruner2018DensitySolids}. The CDW gap can be measured by angle-resolved photoemission spectroscopy (ARPES) \cite{Sobota2021Angle-resolvedMaterials}. On the other hand, from a structural perspective, the CDW order parameter can be measured in a diffraction experiment from the intensity of superlattice peaks \cite{Warren1990X-rayDiffraction}. However, neither equilibrium-state ARPES nor X-ray diffraction (XRD) experiment on $R$Te$_3$ succeeded in unambiguously unveiling the origin of $a$-CDW \cite{Moore2010Fermi/math,Brouet2008,Banerjee2013ChargeTellurides,Ru2008}. 

\begin{figure*}[htb!]
\includegraphics[width=0.8\textwidth]{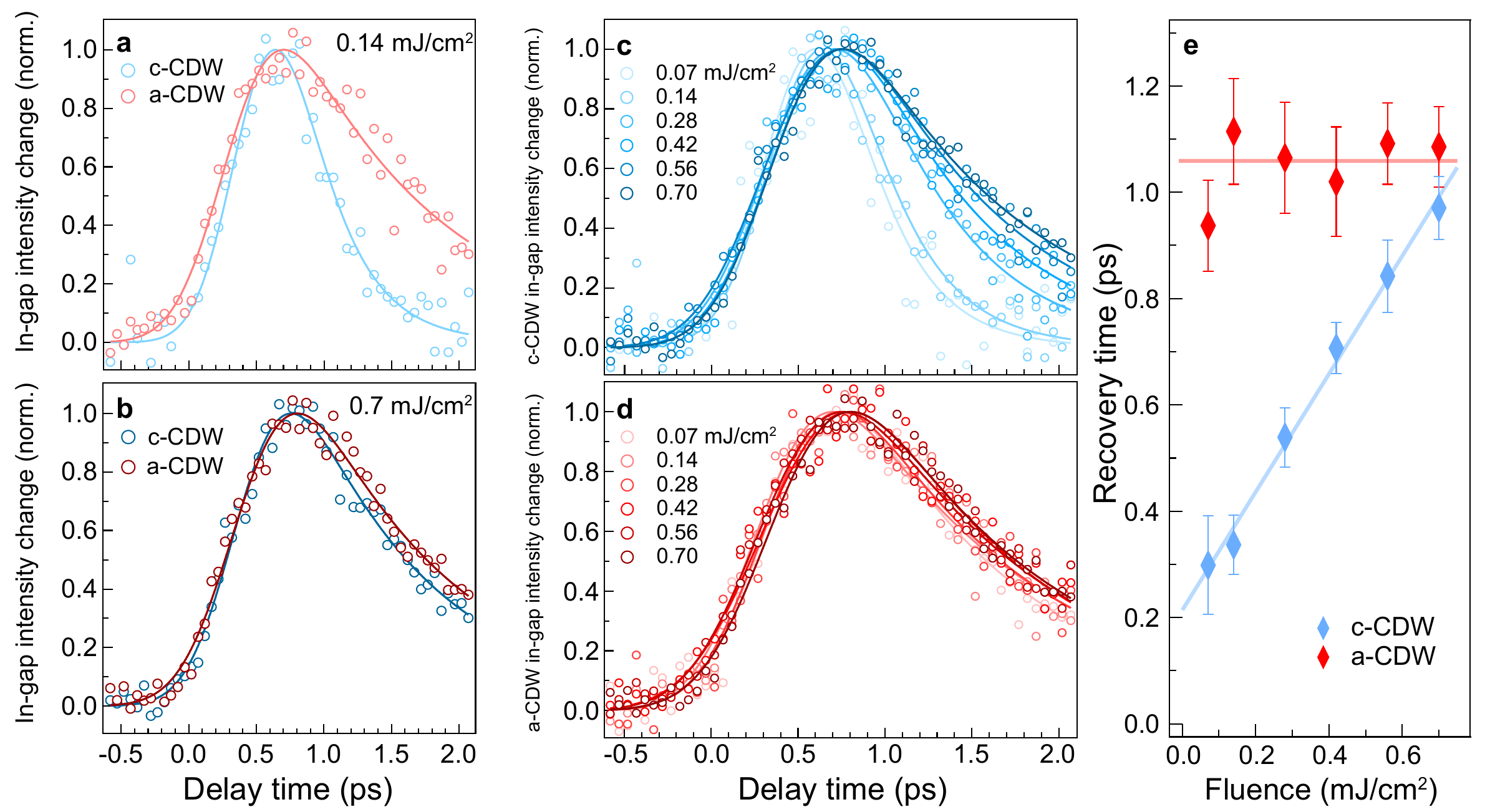}
\caption{\textbf{Distinct photoinduced evolutions of the two CDW gaps in ErTe$_\text{3}$.} \textbf{a,b}~Comparison between $a$-CDW (red) and $c$-CDW (blue) in-gap spectral weight time evolution with low (a) and high (b) excitation fluences. In-gap spectral weights are obtained by integrating over the energy and momentum windows specified in Fig.~\ref{fig:Fig2}a. Spectral weights are normalized between 0 and 1. \textbf{c,d}~Distinct fluence dependence of the spectral weight evolution at $E_F$ in $\Delta_c$(c) and $\Delta_a$(d). \textbf{e}~Characteristic recovery times of the $a$- and $c$-CDW as a function of excitation fluence. The recovery times are obtained by fitting the decaying part of time traces in \textbf{c} and \textbf{d} to an exponential function. The error bars represent 1~s.d. of the fitting errors.}
\label{fig:Fig3}
\end{figure*}

When steady-state techniques fail to capture the physics behind a phase transition, an alternative route is to resort to non-equilibrium measurements. In order to study the CDW phase transitions in $R$Te$_3$ in the non-equilibrium regime and eventually understand the previously unknown driving mechanism of the $a$-CDW---as well as its relation to the $c$-CDW---an ultrafast technique that can simultaneously trace the dynamics of the $a$-CDW and $c$-CDW order parameters is highly desired. To this end, we deploy time-and-angle-resolved photoemission spectroscopy (trARPES). Similar to its steady-state counterpart, trARPES measures the single-electron spectral function \cite{Sobota2021Angle-resolvedMaterials,Lv2019Angle-resolvedMaterials}, which encodes the CDW gap near the Fermi surface. By simultaneously tracking the CDW gap dynamics upon laser excitation, we can compare the different responses of the $a$- and $c$-CDW. We note, however, that a simultaneous measurement of both CDW gaps is not trivial. We not only need a photon energy high enough (>10~eV) to cover enough area in momentum space but also a detection scheme that can resolve both gaps at the same time. A trARPES setup with an angle-resolved time-of-flight (ARTOF) detector with a high harmonic generation (HHG)-based light source, however, can overcome both challenges \cite{Boschini2024Time-resolvedMaterials}.

We, thus, perform trARPES measurements to examine the differences in CDW formation mechanisms in ErTe$_3$, a member of the $R$Te$_3$ family which offers accessibility to both $T_{c1}$ ($\sim265$~K) and $T_{c2}$ ($\sim160$~K). A schematic of the trARPES experiment is illustrated in Fig.~\ref{fig:Fig1}c, where a 10.8~eV extreme-ultraviolet (XUV) pulse was used to probe the electronic structure of ErTe$_3$ via photoemission, following photoexcitation by a 200-fs near-infrared (NIR) laser pulse centered at 1.2~eV. The photoelectrons are collected by an ARTOF analyzer, which, compared to conventional hemispherical analyzers, has the advantage of mapping out the complete three-dimensional (3D) spectrum $I(E,k_x,k_y)$ without rotating the sample or scanning bias voltage \cite{Sobota2021Angle-resolvedMaterials,Lv2019Angle-resolvedMaterials,Lee2020}. This is critical when monitoring two different CDW gaps simultaneously. Fig.~\ref{fig:Fig1}d shows the Fermi surface map from the 10.8~eV ARPES spectrum. The green circle corresponds to the same area in the Brillouin zone as denoted in Fig.~\ref{fig:Fig1}a, with the solid dot marking the orientation. The Fermi surface mapping is labeled by $k_\perp$ and $k_\parallel$, oriented at 45$^\circ$ with respect to the in-plane axes $a$ and $c$. This measurement window allows the simultaneous measurements of both the $c$-CDW gap and $a$-CDW gap. From the Fermi surface, we clearly observe two gapped regions. Fig.~\ref{fig:Fig1}e and \ref{fig:Fig1}f show the energy-momentum dispersion cuts along the blue and red dashed lines in Fig.~\ref{fig:Fig1}d respectively. From the dispersion cuts, we observed the $c$-CDW gap $\Delta_c\approx0.14$~eV and the $a$-CDW gap $\Delta_a\approx0.04$~eV, consistent with previous ARPES measurements \cite{Moore2010Fermi/math,Lee2020}. 

Having the signatures of both CDW orders confirmed, we proceed with pump-probe experiments. We use a 1.2~eV ultrafast laser pulse to excite the system and then track the photoemission spectrum. Figure~\ref{fig:Fig2} shows the snapshots of ARPES spectrum, including Fermi surface map (a) and band dispersion around $a$-CDW (b) and $c$-CDW gaps (c), at different delay time points with respect to the pump pulse. Upon photoexcitation ($t=0$~ps), the conduction bands are populated, while leaving the CDW gaps open at 0.1~ps [Fig.~\ref{fig:Fig2}(a2), (b2), (c2)]. The CDW gaps then gradually collapse, manifesting as spectral weights emerging in the CDW gaps that connect the valence bands and conduction bands [Fig.~\ref{fig:Fig2}(a3)--(a4), (b3)--(b4), (c3)--(c4)]. The CDW gaps eventually reopen on a timescale of around 2~ps [Fig.~\ref{fig:Fig2}(a5), (b5), (c5)] \cite{Zong2021proceeding}. 

\begin{figure*}[htb!]
\includegraphics[width=\textwidth]{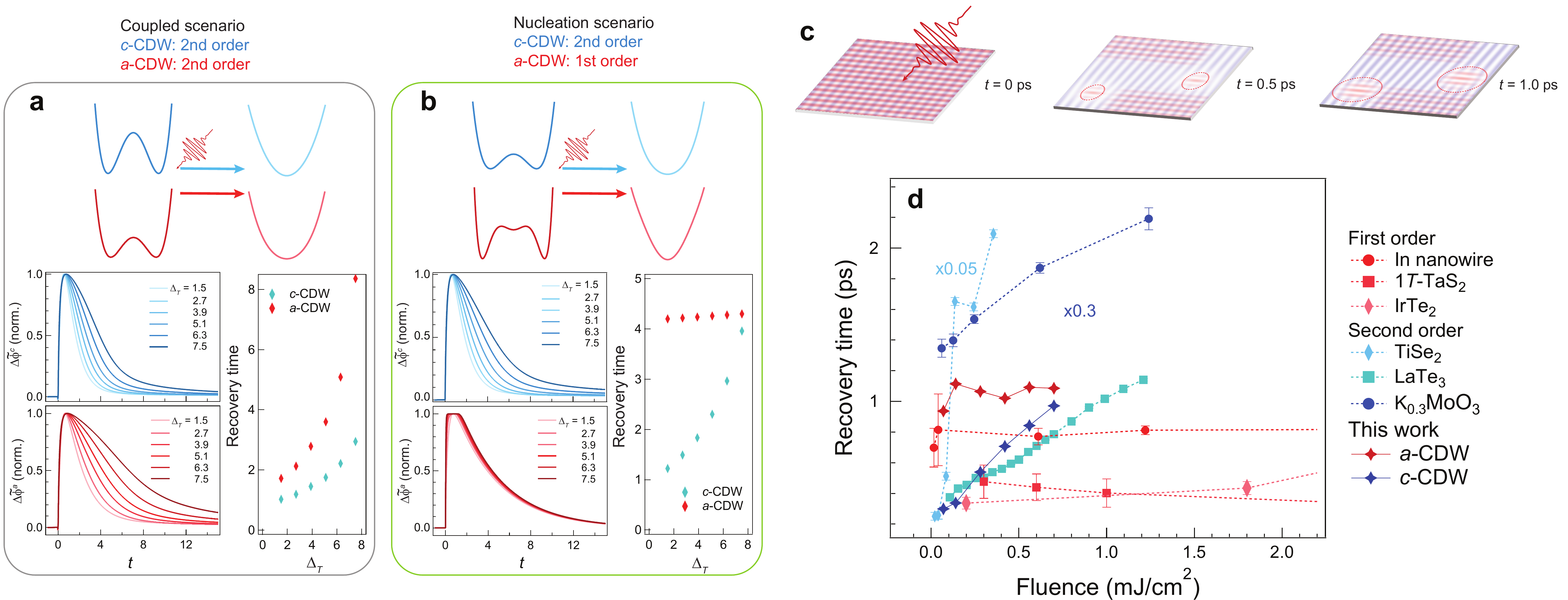}
\caption{\textbf{Nucleation and growth of 
\textit{a}-CDW domains in ErTe$_\text{3}$} \textbf{a,b} Time-dependent Ginzburg-Landau (TDGL) modeling of two scenarios. \textbf{a}~Coupled scenario where the $a$- and $c$-CDW are two competing second-order phase transitions. \textbf{b}~Nucleation scenario where the $a$-CDW is a nucleation-like first-order phase transition. In both panels, we present a qualitative sketch of the free energy landscapes of the $c$- (blue) and $a$-CDW (red). We plot the modeled amplitude dynamics as a function of time $t$ for both order parameters (with a flipped sign) and extract their recovery time as a function of quench strength $\Delta_T$. The nucleation scenario fits the experimental observations significantly better. \textbf{c}~Schematic of nucleation and growth of $a$-CDW after photo-induced quenching of both CDW orders. The dashed ellipses represent the boundary of growing domains. \textbf{d}~The CDW amplitude recovery times as a function of fluence in representative first-order (red) and second-order (blue) CDW phase transitions reported by previous studies\cite{TaS2Mann2016,IrTe2Ideta2018,InHorstmann2020,TiSe2Duan2021,Zong2019b,BBTomeljak2009}. Data of 1$T$-TiSe$_2$ are extracted from trARPES measurements. All other data are extracted from transient reflectivity measurements. The equivalence between the two techniques in the context of measuring order parameter amplitude recovery was shown in ref.~\cite{Zong2019b}.}
\label{fig:Fig4}
\end{figure*}

To understand the difference between the $a$- and $c$-CDW, we investigate CDW relaxation dynamics after photoexcitation. The recovery timescale of CDWs contains rich information about the nature of light-induced CDW excitation and phase transition, such as the formation of topological defects and phase fluctuations \cite{Zong2019,Kogar2020,Sun2020}. An important tuning knob is the excitation fluence, which can typically unveil criticality and scaling laws in light-induced dynamics, especially in $R$Te$_3$ \cite{Zong2019b,Orenstein2023,Trigo2021}. We, thus, perform continuous time scans at different excitation fluences. We plot the spectral weight in the respective CDW gaps as a function of the pump-probe delay to capture the recovery dynamics upon photoexcitation, similar to the approach adopted in previous studies \cite{Zong2019,Maklar2021NonequilibriumLimit,Lee2020}. Analyzing the recovery dynamics of the $a$- and $c$-CDW with a low pump fluence (0.14~mJ/cm$^2$, plotted in Fig.~\ref{fig:Fig3}a), the recovery of the $a$-CDW is significantly slower than the $c$-CDW. On the contrary, when pumping with a high fluence (0.7~mJ/cm$^2$, plotted in Fig.~\ref{fig:Fig3}b), this difference in the recovery dynamics is significantly smaller. Comparing time scans with a series of fluences, we find a striking difference in the relaxation dynamics as a function of fluence. The recovery time scales with fluence in $c$-CDW (Fig.~\ref{fig:Fig3}c). This is quantitatively consistent with our previous observation of a recovery that is slowed down by topological defects in LaTe$_3$, where only $c$-CDW is present \cite{Zong2019} (see also Fig.~S12 in the Supplemental Information). On the other hand, the recovery dynamics of the $a$-CDW (Fig.~\ref{fig:Fig3}d), in stark contrast, barely show any variation across the fluence range that spans an order of magnitude. We first note that the fluence independence is not due to the saturated CDW gap closure, as we can clearly observe different leading edges of CDW gaps in the energy distribution curves (EDCs) under different fluences (see Supplemental Information Sec.~I.B). To qualitatively compare the recovery dynamics, we fit the time traces to an error function, which phenomenologically models the initial CDW melting, multiplied by an exponential decay \cite{ZongThesis}. The time constants extracted from exponential fitting as a function of fluence are plotted in Fig.~\ref{fig:Fig3}e (ref.~\cite{ZongThesis}). We observe a nearly constant recovery time constant for the $a$-CDW and a linear function of fluence for recovery time constants in $c$-CDW, approaching that of $a$-CDW at high fluence. 

The stark contrast in recovery dynamics indicates a fundamental difference in the formation mechanisms between the $c$- and $a$-CDW. Recall that the formation of the $a$-CDW is not accompanied by a soft phonon \cite{Maschek2018Competing/math}. One possible explanation for the missing soft phonon is a nucleation and growth of the CDW domains \cite{Aubry1992}. This scenario may also be responsible for the nearly fluence-invariant recovery dynamics in the $a$-CDW as the recovery rate of CDW in this case is set by the nucleation rate. 

To examine this hypothesis, we perform a time-dependent Ginzburg-Landau (TDGL) modeling. We first note that the slowing down of $c$-CDW recovery rate with fluence is consistent with previous works \cite{Zong2019,Maklar2022CoherentSystem}. Combined with its well-studied behaviors of stereotypical second-order phase transition in equilibrium \cite{Moore2010Fermi/math,Maschek2015Wave-vector-dependentE3,Maschek2018Competing/math}, we can thus model the $c$-CDW as a second-order phase transition. With this baseline set for the modeling, we compare two different scenarios. In the first scenario, we assume a coupled scenario where $a$-CDW is another second-order phase transition competing with the $c$-CDW: the $a$-CDW is essentially the $c$-CDW counterpart in the perpendicular $a$-axis that competes for electrons near the Fermi level and lost to the $c$-CDW at $T_{c1}$. This is a scenario suggested by several previous studies on $R$Te$_3$ systems \cite{Straquadine2022Evidencemrow,Kivelson2023EmergentPoint,Banerjee2013ChargeTellurides}. The second scenario models the $a$-CDW as a nucleation-like first-order phase transition. We expect this scenario to remain valid even if a weak coupling between the order parameters is introduced.

A detailed description of the TDGL model can be found in the Supplemental Information. In both cases, the two CDW orders can be represented by a complex scalar field $\Phi^c = \phi^c_{1} + i \phi^c_{2}$ and $\Phi^a=\phi^a_{1}+ i \phi^a_{2}$. To directly compare with the trARPES data presented in Fig.~\ref{fig:Fig3}, we calculate the dynamics of the amplitudes of both scalar fields with a flipped sign $\Tilde{\phi}^{o}=-\sqrt{ {(\phi^{o}_{1}})^2+({\phi^{o}_{2}})^2}$, where $o=a \text{ or } c$, all normalized between 0 and 1. As shown in Fig.~\ref{fig:Fig4}a, in the coupled scenario, the recovery time of both CDW orders slows down upon an increase in the quench strength $\Delta_T$ due to the photo-induced fluctuations, as reported by previous studies \cite{Zong2019,Kogar2020,Zong2021RoleOrder,Dolgirev2020Self-similarExperiments}. This behavior significantly deviates from the dynamics observed in our trARPES experiments, where the recovery timescale of the $a$-CDW is nearly fluence-independent. On the contrary, the nucleation scenario presented in Fig.~\ref{fig:Fig4}b qualitatively reproduces the different recovery dynamics of the $a$- and $c$-CDW, indicating a possible nucleation and growth origin of the $a$-CDW. 

The observation that the recovery time of the $a$-CDW is nearly independent of the quench strength can be understood in an intuitive physical picture. Upon photoexcitation, some domains of the $a$-CDW are suppressed. Each domain can be defined with respect to a defect in the center of the domain. These defects function as seeds for the nucleation of the $a$-CDW. Upon cooling of the electronic subsystem, a new minimum with lower energy forms. The height of the local free energy barrier determines the probability of nucleating into the new minima. Once the local nucleation happens, the mean-field value is suddenly large and fluctuations become negligible. The bubble-like CDW domain then grows spherically like an inflating bubble at the rate of nucleation until it impinges another growing bubble (see Fig.~\ref{fig:Fig4}c and Supplemental Information Sec.~II.B.2). The recovery rate of the $a$-CDW is then governed by the nucleation rate set by the pressure difference, arising from the difference in free energy, between the $a$-CDW phase and the phase without the $a$-CDW order, independent of light-induced fluctuations. An increase in laser fluence or, equivalently, the quench strength increases the number of nucleation bubbles but not the recovery rates. The recovery time of $a$-CDW is thus barely fluence-dependent. 

The TDGL theory and the physical picture described above should, in principle, be generally applicable to other systems beyond ErTe$_3$. Comparing previous pump-probe experiments that measured the order parameter amplitude of some prototypical CDWs, as shown in Fig.~\ref{fig:Fig4}d, we find that in material systems with first-order CDW phase transitions, such as 1$T$-TaS$_2$, IrTe$_2$, and indium nanowires on silicon substrates \cite{TaS2Mann2016,IrTe2Ideta2018,InHorstmann2020}, the recovery timescales are insensitive to the laser fluence in the fluence regime we studied in this work. While in systems with second-order CDW transitions, such as LaTe$_3$, 1$T$-TiSe$_2$, and K$_{0.3}$MoO$_3$ (blue bronze) \cite{Zong2019b,TiSe2Duan2021,BBTomeljak2009}, the recovery slows down with growing fluence due to light-induced fluctuations (see Sec.~III in Supplemental Information for details). From this perspective, ErTe$_3$ provides a unique platform as a controlled comparison between CDW transitions of different origins, as marked in Fig.~\ref{fig:Fig4}d with red and blue diamonds. However, we notice that exceptions to this simple time-domain classification may occur in some transient reflectivity measurements where the time evolution of reflectivity involve significant contributions from other physical processes \cite{Venturini2023}. This, on the other hand, also demonstrates the advantage of trARPES measurements in isolating the dynamics of the order parameter. This comparative study, though not exhaustive, supports the validity of our time-domain protocol based on TDGL theory.

We would like to highlight that the combination of experiment, analysis, and modeling presented above demonstrates the potential of using non-equilibrium techniques to identify the true nature of a phase transition. Measuring a first-order phase transition in an equilibrium-state experiment without nano-scale spatial resolution typically involves averaging over many local first-order phase transitions. This averaging process, together with contributions from disorder, may give rise to a smoothly varying order parameter without hysteresis, resembling a second-order phase transition. A non-equilibrium technique offers an alternative route for differentiating the nature of phase transitions via vastly different recovery dynamics of distinct orders when driven out of equilibrium. 

The unique capability of ARTOF-based trARPES measurements enabled us to simultaneously track the time evolution of both CDW order parameters in ErTe$_3$ upon photoexcitation. Owing to the simple ARPES measurement geometry without sample rotation, we conclusively observed distinct recovery rates of the $a$- and $c$-CDW under a series of laser fluences. The ultrafast pump-probe measurement allows the simultaneous observation of the $a$- and $c$-CDW formation, which is not possible in steady-state measurements due to the distinct transition temperatures. The fluence-independent recovery time of the $a$-CDW suggests a nucleation and growth behavior, which is further confirmed by our TDGL model. Extending this result to infinitesimal fluence where the excitation is close to an equilibrium perturbation and the distinction between the $a$- and $c$-CDW is most pronounced, we infer that the $a$-CDW is more likely formed through a first-order phase transition in equilibrium. This result potentially answers the long-standing debate on the nature of $a$-CDW phase transition in $R$Te$_3$ systems. Furthermore, the pump-probe-based protocol developed in this work could be applied to other physical systems with multiply intertwined phases beyond CDWs to identify phase nucleation. This non-equilibrium approach enables the classification of phase transitions in novel quantum materials, whose nature is challenging to identify using typical experimental techniques in the equilibrium regime. 

\section{Materials and Methods}
\subsection{Material preparation}
Single crystals were grown via a self-flux technique. Elements in the molar ratio $R_x$Te$_{1-x}$, $x=0.015$--0.003, were put into alumina crucibles and vacuum sealed in quartz tubes. The mixtures were heated to 800–900$^\circ$C and slowly cooled to end temperatures in the range of 500–600$^\circ$C. The remaining melt was decanted in a centrifuge. Before trARPES measurements, bulk single crystals were cleaved in ultrahigh vacuum ($<10^{-10}$~Torr) to expose a pristine surface.

\subsection{Experimental setup of trARPES}
The 1030~nm (1.2~eV) output of a commercial Yb-fiber laser system (Tangerine, Amplitude) operating at 300~kHz was split into pump and probe branches. The pump branch was focused onto the cleaved surface of ErTe$_3$ to excite the sample. The probe branch was first frequency-tripled to 344~nm (3.6~eV) and then focused onto Ar gas ejected by a gas-jet nozzle to generate the 9th harmonic at 114~nm (10.8~eV). The resulting extreme-ultraviolet (XUV) pulse was passed through a custom-built time-preserving grating monochromator (McPherson OP-XCT) to minimize pulse width broadening and enhance efficiency. After exiting the monochromator, the XUV pulse was focused onto the sample surface by a toroidal mirror. Photoelectrons were collected by a time-of-flight detector (Scienta ARTOF 10k), which made simultaneous measurements of the energy and in-plane momenta possible. 

\section{Acknowledgements}
The authors thank Mariano~Trigo, Gal~Orenstein, Hugo~Wang, So-yung~Kim, Anshul~Kogar, Francesco Grandi, Jonathan Curtis, and Minhui~Zhu for stimulating discussions. We thank Xinyue Lu for the help with the artistic rendering of the figures. The work at MIT was supported by the U.S. Department of Energy, National Science Foundation, BES DMSE, and Gordon and Betty Moore Foundation's EPiQS Initiative grant GBMF9459 (instrumentation). B.L. acknowledges support from the Ministry of Science and Technology of China (2023YFA1407400), the National Natural Science Foundation of China (12374063), the Shanghai Natural Science Fund for Original Exploration Program (23ZR1479900).

\end{document}